\documentclass[a4paper,10pt]{article}

\usepackage{amssymb}
\usepackage{amsmath}

\newcommand{\beq}{\begin{equation}}
\newcommand{\eeq}{\end{equation}}
\newcommand{\nec}{\newcommand}
\nec{\cts}{conformal transformations }
\nec{\fourg}{g_{\alpha\beta}} 
\nec{\grad}{\bigtriangledown}
\nec{\fourr}{^{(4)}R}
\nec{\detg}{^{(4)}g}
\nec{\eins}{\Biggl(R_{\alpha\beta} - \frac{1}{2}g_{\alpha\beta}R\Biggr)}
\nec{\kk}{K^{ab}K_{ab}}
\nec{\bec}{\begin{center}}
\nec{\eec}{\end{center}}
\nec{\bb}{B^{ab}B_{ab}}
\nec{\rp}{R - 8\frac{\grad^{2}\psi}{\psi}}
\nec{\pipi}{\pi^{ab}\pi_{ab}}
\nec{\beqq}{\begin{equation*}}
\nec{\eeqq}{\end{equation*}}
\nec{\V}{V(\psi)}

\begin{document}

\baselineskip=.65cm
\begin{center}
\noindent{\LARGE\bf{Scale-invariant gravity: Spacetime recovered}}

\noindent{Bryan Kelleher}
\\
\noindent{\emph{Physics Department, University College, Cork, Ireland}}
\\
\noindent{E-mail: bk@physics.ucc.ie}
\end{center}
\renewcommand{\abstractname}{}
\begin{abstract}
\noindent{The configuration space of general relativity is \emph{superspace} - 
the space of all Riemannian $3$-metrics modulo diffeomorphisms. However, it 
has been argued that the configuration space for gravity should be 
\emph{conformal superspace} - the space of all Riemannian $3$-metrics modulo
diffeomorphisms \emph{and} conformal transformations. Recently a manifestly 
$3$-dimensional theory was constructed with conformal superspace as the 
configuration space. Here a fully $4$-dimensional action is constructed so as 
to be invariant under conformal transformations of the $4$-metric using general
 relativity as a guide. This action is then decomposed to a $(3+1)$-dimensional
 form and from this to its Jacobi form. The surprising thing is that the new 
theory turns out to be \emph{precisely} the original $3$-dimensional theory. 
The physical data is identified and used to find the physical representation of
 the theory. In this representation the theory is extremely similar to general 
relativity. The clarity of the $4$-dimensional picture should prove very useful
 for comparing the theory with those aspects of general relativity which are 
usually treated in the $4$-dimensional framework.}
\end{abstract}

\section{Introduction}
As formulated by Einstein, the natural arena for gravity as represented by 
general relativity is spacetime. We have a purely $4$-dimensional structure
and the $4$-geometry reigns. The reformulation of the theory in canonical 
dynamical form by Dirac \cite{dir} and Arnowitt, Deser and Misner (ADM) 
\cite{adm} led away from the $4$-dimensional picture and placed the emphasis
more on the $3$-geometry. The configuration space is superspace and general 
relativity describes the evolution of the $3$-geometry in time 
(geometrodynamics). York \cite{yor} went further and the identified the 
conformal $3$-geometry with the dynamical degrees of freedom of the 
gravitational field. The correct configuration space for gravity should not be
superspace but rather \emph{conformal superspace} - superspace modulo 
conformal transformations.
\\ \\
Alas, general relativity and conformal superspace are not entirely compatible.
Barbour and \'{O} Murchadha \cite{bom} constructed a theory with conformal 
superspace at the very core. They took the Jacobi action for general relativity
of Baerlein, Sharp and Wheeler (BSW) \cite{bsw} and constructed an action which
is invariant under conformal transformations of the $3$-metric. Conformal 
superspace arises naturally in this theory.
\\ \\
In this paper we construct a \emph{four}-dimensional action based on conformal
transformations of the \emph{four}-metric. We then decompose this to a 
$(3 + 1)$-dimensional form and from this we find the Jacobi action of the 
theory. Incredibly, it turns out to be the same as that of Barbour and \'{O} 
Murchadha. We will begin with a review of some general relativity before 
considering the new theory.
\section{The Action}  
The Einstein-Hilbert action of general relativity is well known. It has 
the form
\beq \label{eh} S = \int \sqrt{-^{(4)}g}\: \fourr \: d^{4}x \eeq 	
where $g_{\alpha\beta}$ is the 4-metric and $^{(4)}R$ is the four-dimensional 
Ricci scalar. The action is varied with respect to $g_{\alpha\beta}$ and the 
resulting equations are the (vacuum) Einstein equations
\beq   G^{\alpha\beta} = \Biggl(R^{\alpha\beta} - 
\frac{1}{2}g^{\alpha\beta}R\Biggr) = 0   \eeq
We would like to construct an action which is invariant under \cts of the 
metric
\beq \label{ct1} \fourg \longrightarrow \Omega^{2}\fourg \eeq
where $\Omega$ is a strictly positive function using the Einstein-Hilbert 
action as a guide. First we need to develop some machinery for dealing with 
conformal transformations.
\subsection{Dimensional Properties of Conformal Transformations}
A supposed problem with conformal transformations and different numbers of
dimensions is that various coefficients change when the number of dimensions
changes. This turns out not to be a problem in this analysis as will be shown.
\\ \\
Let us consider conformal transformations and the scalar curvature. If we make
a conformal transformation of the metric of the form in equation (\ref{ct1}) 
above then the Ricci tensor transforms as \cite{wald}
\beq
\begin{split} ^{(n)}R_{\alpha\beta} \longrightarrow &\:^{(n)}R_{\alpha\beta} 
+ 2(n - 2)\frac{(\grad_{\alpha}\Omega)\grad_{\beta}\Omega}{\Omega^{2}}
 - (n - 2) \frac{\grad_{\alpha}\grad_{\beta}\Omega}{\Omega} \\
&+ (3 - n)g_{\alpha\beta}\frac{(\grad_{\gamma}\Omega)\grad^{\gamma}\Omega}
{\Omega^{2}}
- g_{\alpha\beta}\frac{\grad_{\gamma}\grad^{\gamma}\Omega}{\Omega} \end{split}
\eeq
Then we can contract to find that the scalar curvature transforms as
\beq ^{(n)}R \longrightarrow \Omega^{-2}\biggl(\, ^{(n)}R - 
2(n-1)g^{\alpha\beta}\frac{\grad_{\alpha}\grad_{\beta}\Omega}{\Omega} + 
(n-1)(4-n)\frac{\grad_{\gamma}\Omega\grad^{\gamma}\Omega}{\Omega^{2}}\, 
\biggr) \eeq
where $n$ is the number of dimensions. A consequence is that the combination
\beq \phi^{2/s}\biggl(\, ^{(n)}R - 
\frac{4(n-1)}{(n-2)}g^{\alpha\beta}\frac{\grad_{\alpha} 
\grad_{\beta}\phi}{\phi}\,\biggr) \eeq 
is conformally invariant for any scalar function $\phi$ under the combined 
transformation
\beq g_{\alpha\beta} \longrightarrow \Omega^{2}g_{\alpha\beta} \: \: \: 
\text{,} \: \: \: \phi \longrightarrow \Omega^{s}\phi \eeq
where $s = 1 - \frac{n}{2}$. While this is true in any number of dimensions we
are of course most concerned with the $3$-dimensional and $4$-dimensional 
cases. In $3$ dimensions we have $s=-\frac{1}{2}$. Thus we get that
\beq \phi^{-4}\biggl(\; ^{(3)}R - \frac{8\grad^{2}\phi}{\phi}\;\biggr) \eeq
is conformally invariant under the transformation 
\beq g_{ab} \longrightarrow 
\Omega^{2}g_{ab} \:\:\: \text{,} \:\:\: \phi \longrightarrow 
\frac{\phi}{\sqrt{\Omega}} \eeq
In four dimensions $s = -1$ and the combination
\beq \phi^{-2}\biggl(\,\fourr - \frac{6\Box\phi}{\phi}\,\biggr) \eeq
is conformally invariant under the transformation \beq g_{\alpha\beta} 
\longrightarrow \Omega^{2}g_{\alpha\beta} \:\:\: \text{,} \:\:\: 
\phi\longrightarrow \frac {\phi}{\Omega} \eeq 
Then the combination
\beq \sqrt{-^{(4)}g}\phi^{2}\biggl(\,\fourr - \frac{6\Box\phi}{\phi}\,\biggr) 
\eeq is also conformally invariant.
This will be our Lagrangian density $\mathcal{L}$. Thus our action is
\beq  S = \int\mathcal{L}\: d^{4}x  \eeq
Before we decompose this to a $(3 + 1)$-dimensional form let us consider the 
$4$-dimensional structure and see what emerges.
\subsection{Varying with respect to $\fourg$}
The variation with respect to $\fourg$ is quite straightforward. The resulting
equations of motion are
\beq -\phi^{2}\biggl(R^{\alpha\beta} - \frac{1}{2}g^{\alpha\beta}R\biggr) 
+ 4 \grad^{\alpha}\phi\grad^{\beta}\phi - 
g^{\alpha\beta}\grad_{\gamma}\phi\grad^{\gamma}\phi - 2 \phi \grad^{\alpha} 
\grad^{\beta}\phi + 2 g^{\alpha\beta}\phi \Box \phi = 0  \eeq
This looks quite complicated but it is actually just
\beq \label{ceins} \overline{G^{\alpha\beta}} = 0 \eeq where 
$\overline{G^{\alpha\beta}}$ is just the Einstein tensor conformally 
transformed with conformal factor $\phi$. Equivalently, this is the Einstein 
tensor for the metric $\phi^{2}\fourg$. This interpretation will prove useful 
later.
\subsection{Varying with respect to $\phi$}
Again, this variation is fairly straightforward. We get
\beq \label{phi}\fourr  - \frac{6\Box\phi}{\phi} = 0\eeq
This is actually the trace of (\ref{ceins}) \footnote{We use the signature 
$\{-,+,+,+\}$ which results in the minus sign here.} and so, as such, is 
redundant. This can be viewed as a result of $\phi$ being pure gauge. Work by 
Barbour on the variation of gauge variables \cite{jb} shows that for a pure 
gauge variable $\psi$ we may vary the action with respect to both $\psi$ and 
its time derivative $\dot{\psi}$ independently. We are permitted to perform 
so-called free-end point variations. Because $\phi$ is pure gauge here we may 
vary the action with respect to $\phi$ and $\dot{\phi}$ independently. This 
will be crucial in the theory.. We shall return to this.
\subsection{A note on the action}
The form of the action as it stands is not conventional as it contains second 
time derivatives of the metric. However, the combination
\beq \fourr + 2A^{\alpha}_{\;\; ;\alpha}  \eeq
where $A^{\alpha} = \bigl(n^{\alpha}trK + a^{\alpha}\bigr)$ , $n^{\alpha}$ is 
the unit timelike normal and $a^{\alpha}$ is the four-acceleration of an 
observer travelling along $\mathbf{\underline{n}}$, contains no second time 
derivatives. (The coordinates $\alpha$ are general.) We write our Lagrangian as
\beq \mathcal{L} = \sqrt{-^{(4)}g}\phi^{2}
\biggl(\; \fourr + 2A^{\alpha}_{\;\; ;\alpha} - 2A^{\alpha}_{\;\; ;\alpha}
 - \frac{6\Box\phi}{\phi}\; \biggr) \eeq 
which then becomes
\beq \mathcal{L} = \sqrt{-^{(4)}g}
\biggl(\phi^{2}\biggl(\; \fourr + 2A^{\alpha}_{\;\; ;\alpha}\;\biggr) + 
4\phi\phi_{,\alpha}A^{\alpha} + 
6g^{\mu\nu}\grad_{\mu}\phi\grad_{\nu}\phi\biggr) \eeq
after some integration by parts. 
\\ \\
This Lagrangian contains no second time derivatives of the metric. Varying this
with respect to $\phi$ and $\dot{\phi}$ gives two conditions which combine to
give equation (\ref{phi}). Although we may do these variations here in a 
general coordinate form it will be more instructive to do a $(3+1)$-dimensional
decomposition and get the corresponding equations there.
\section{(3+1)-Decomposition}
Before we consider the new theory it will be instructive to recall the ADM 
treatment of general relativity as much of this will carry straight over to the
new theory.
\\ \\
The idea in the ADM treatment is that a thin-sandwich $4$-geometry is 
constructed from two $3$-geometries separated by the proper time $d\tau$. The 
$4$-metric found from the ADM construction is
\begin{gather}
\begin{Vmatrix} ^{(4)}g_{00} & ^{(4)}g_{0k} \\ \\  ^{(4)}g_{i0} & ^{(4)}g_{ik} 
\end{Vmatrix} \quad
= \quad \begin{Vmatrix} (N^{s}N_{s} - N^{2}) & N_{k} \\ \\  N_{i} & g_{ik}
\end{Vmatrix}
\end{gather}
$N = N(t,x,y,z)$ is the lapse function given by
\beq d\tau = N(t,x,y,z)dt \eeq
and $N^{i}= N^{i}(t,x,y,z)$ are the shift functions given by
\beq x^{i}_{2}(x^{m}) = x^{i}_{1} - N^{i}(t,x,y,z)dt \eeq
where $x^{i}_{2}$ is the position on the ``later'' hypersurface corresponding 
to the position $x^{i}_{1}$ on the ``earlier'' hypersurface. The indices in the
 shift are raised and lowered by the $3$-metric $g_{ij}$.
\\ \\
The reciprocal $4$-metric is
\begin{gather}
\begin{Vmatrix} ^{(4)}g^{00} & ^{(4)}g^{0k} \\ \\  ^{(4)}g^{i0} & ^{(4)}g^{ik} 
\end{Vmatrix} \quad
= \quad \begin{Vmatrix}  -1/N^{2} & N^{k}/N^{2} \\ \\  N^{i}/N^{2} & 
g^{ik} - N^{i}N^{k}/N^{2}
\end{Vmatrix}
\end{gather}
The volume element has the form
\beq \sqrt{^{(4)}g}\;d^{4}x = N\sqrt{g}\;dt\;d^{3}x \eeq
This construction of the $4$-metric also automatically determines the 
components of the unit timelike normal vector $\mathbf{\underline{n}}$. We get
\beq n_{\beta} = (-N,0,0,0) \eeq
and raising the indices using $^{(4)}g^{\alpha\beta}$ gives us
\beq \label{un} n^{\alpha} = (1/N, -N^{m}/N) \eeq
\\
Consider now the Einstein-Hilbert action
\beq S = \int \sqrt{-^{(4)}g}\: \fourr \: d^{4}x \eeq
Using the Gauss-Codazzi relations we get
\beq \fourr = R - (trK)^{2} + \kk - 2A^{\alpha}_{\;\; ;\alpha} \eeq 
where $A^{\alpha}$ is given by (as earlier)
\beq A^{\alpha} = \bigl(n^{\alpha}trK + a^{\alpha}\bigr) \eeq
$n^{\alpha}$ is the unit timelike normal and 
\beq a^{\alpha} = n^{\alpha}_{\;\; ;\beta}n^{\beta} \eeq
is the four-acceleration of an observer travelling along 
$\mathbf{\underline{n}}$. It is easily verified that $a^{0} = 0$ and that 
$a^{i} = \frac{\grad^{i}N}{N}$. Substituting into the action gives
\beq \label{admact} S = \int N\sqrt{g}(R - (trK)^{2} + \kk) dtd^{3}x \eeq
where the total divergence $A^{\alpha}_{\;\; ;\alpha}$ has been discarded. 
$\mathbf{K}$ is the \emph{extrinsic curvature} given by
\beq \mathbf{K} = -\frac{1}{2}\mathbf{\pounds_{\underline{n}}}\mathbf{g} \eeq
the Lie derivative of the $3$-metric  $\mathbf{g}$ along 
$\mathbf{\underline{n}}$. In the coordinates we are using here the extrinsic 
curvature takes the form
\beq K_{ab} = -\frac{1}{2N}\biggl(\frac{\partial g_{ab}}{\partial t} - 
N_{a:b} - N_{b;a} \biggr) \eeq
The action is varied with respect to $\frac{\partial g_{ab}}{\partial t}$ to 
get the canonical momentum
\beq \pi^{ab} = \sqrt{g}\bigl(g^{ab}trK - K^{ab}\bigr) \eeq
and varied with respect to $N$ and $N_{a}$ to give the initial value equations
\beq \mathcal{H} = 0 \:\:\: \text{and} \:\:\: \mathcal{H}^{a} = 0 \eeq
respectively, where
\beq \mathcal{H} = \sqrt{g}\biggl(\pi^{ab}\pi_{ab} - 
\frac{1}{2}(tr\pi)^{2}\biggr) - \sqrt{g}R \eeq
and 
\beq  \mathcal{H}^{a} = -2\pi^{ab}_{\;\;\;\; ;b} \eeq
\\
We are now ready to consider the new action. This is
\beq S = \int \sqrt{-^{(4)}g}\phi^{2}\biggl(\,\fourr - 
\frac{6\Box\phi}{\phi}\,\biggr)d^{4}x \eeq
The $4$-dimensional scalar curvature decomposes as earlier. The action becomes
\beq S = \int \sqrt{-^{(4)}g}\phi^{2}\biggl(R - (trK)^{2} + \kk 
- 2A^{\alpha}_{\;\; ;\alpha} - 6\frac{\Box\phi}{\phi}\biggr) \; d^{4}x \eeq
Let's separate this into two terms $ S_{1} $ and $ S_{2} $ where,
\beq S_{1} = \int \sqrt{-^{(4)}g}\phi^{2}\biggl(R - (trK)^{2} + \kk 
- 2A^{\alpha}_{\;\; ;\alpha}\biggr) \; d^{4}x \eeq
and
\beq \label{s2} S_{2} = - \int 6\sqrt{-^{(4)}g}\phi\Box\phi d^{4}x \eeq
Consider the first term. In the ADM theory $A^{\alpha}_{\;\; ;\alpha}$ leads to
a total divergence which is discarded. However, the presence of the $\phi^{2}$
here changes this. Integrating by parts we get 
\beq -2\phi^{2}A^{\alpha}_{\;\; ;\alpha} \longrightarrow 
2(\phi^{2})_{;\alpha}A^{\alpha} \eeq
discarding the total divergence again. Decomposing this gives
\beq
\begin{split} & 2\Biggl(\dot{\phi^{2}}\biggl(n^{0}trK + a^{0}\biggr) + 
\biggl(\phi^{2}\biggr)_{,i}\biggl(n^{i}trK + a^{i}\biggr)\Biggr) \\
= \;\; & 4\phi\biggl(\dot{\phi}n^{0}trK + \phi_{,i}n^{i}trK\biggr) + 
4\phi\phi_{,i}a^{i} \\
= \;\; & \frac{4\phi}{N}\biggl(\dot{\phi} - \phi_{,i}N^{i}\biggr)trK + 
4\phi\phi_{,i}a^{i} \end{split} \eeq
using equation (\ref{un}). Then,
\beq
\begin{split} 
S_{1} = \;\; & \int N\sqrt{g}\phi^{2}\biggl(R - (trK)^{2} + \kk\biggr)\: dt\: 
d^{3}x \\
 + & \int 4\sqrt{g}\phi\Biggl[\biggl(\dot{\phi} - \phi_{,i}N^{i}\biggr)trK + 
N\phi_{,i}a^{i}\Biggr]\: dt\: d^{3}x \end{split}
\eeq
We must now deal with $S_{2}$. After a little integration by parts this is
\beq S_{2} = \int 6\sqrt{-^{(4)}g}\,g^{\mu\nu}\grad_{\mu}\phi\grad_{\nu}\phi 
d^{4}x \eeq
Decomposing this gives
\beq S_{2} =  \int 6N\sqrt{g} 
\Biggl(-\frac{1}{N^{2}}\biggl(\dot{\phi}\biggr)^{2} + 
\frac{2N^{i}}{N^{2}}\dot{\phi}\phi_{,i} + \biggl(g^{ij} - 
\frac{N^{i}N^{j}}{N^{2}}\biggr)\phi_{,i}\phi_{,j}\Biggr)\: dt\: d^{3}x \eeq
The full action is now
\beq
\begin{split} 
S = \;\; & \int N\sqrt{g}\phi^{2}\biggl(R - (trK)^{2} + \kk\biggr)\: dt\: 
d^{3}x \\
+ & \int 4\sqrt{g}\phi\Biggl[\biggl(\dot{\phi} - \phi_{,i}N^{i}\biggr)trK
 + N\phi_{,i}a^{i}\Biggr]\: dt\: d^{3}x  \\
+ & \int 6N\sqrt{g} \Biggl(-\frac{1}{N^{2}}\biggl(\dot{\phi}\biggr)^{2} 
+ \frac{2N^{i}}{N^{2}}\dot{\phi}\phi_{,i} + \biggl(g^{ij} - 
\frac{N^{i}N^{j}}{N^{2}}\biggr)\phi_{,i}\phi_{,j}\Biggr)\: dt\: d^{3}x 
\end{split}
\eeq
This looks like a much more complicated object than we began with. There will,
however, be much simplification. First, let's write it as
\beq
\begin{split} 
S = \;\; & 
\int N\sqrt{g}\phi^{2}\biggl(R - (trK)^{2} + \kk\biggr)\: dt\: d^{3}x \\
+ & \int 4\sqrt{g}\phi\Biggl[\biggl(\dot{\phi} - \phi_{,i}N^{i}\biggr)trK
 + \grad_{i}\phi\grad^{i}N\Biggr]\: dt\: d^{3}x \\
- & \int \frac{6}{N}\sqrt{g}\biggl(\dot{\phi} - 
\phi_{,i}N^{i}\biggr)^{2}\: dt\: d^{3}x  + 
\int 6N\sqrt{g}\grad_{i}\phi\grad^{i}\phi \: dt\: d^{3}x \end{split}
\eeq
where we have used $ a^{i} = \frac{\grad^{i}N}{N} $. If we set 
$ \theta = -\frac{2}{\phi}\biggl(\dot{\phi} - \phi_{,i}N^{i}\biggr) $
then we get
\beq
\begin{split} 
S = \;\; & 
\int N\sqrt{g}\phi^{2}\biggl(R - (trK)^{2} + \kk\biggr)\: dt\: d^{3}x \\
- & \int 2\sqrt{g}\theta\phi^{2}trK\: dt\: d^{3}x - 
\int \frac{3}{2}\frac{\sqrt{g}\theta^{2}\phi^{2}}{N}\: dt\: d^{3}x \\
+ & \int 6N\sqrt{g}\grad_{i}\phi\grad^{i}\phi \: dt\: d^{3}x + 
\int 4\sqrt{g}\phi\grad_{i}\phi\grad^{i}N\: dt\: d^{3}x \end{split}
\eeq 
This becomes
\beq
\begin{split}
S = \;\; & 
\int N\sqrt{g}\phi^{2}\biggl(R - (trK)^{2} + \kk\biggr)\: dt\: d^{3}x \\
- & \int 2\sqrt{g}\theta\phi^{2}trK\: dt\: d^{3}x - 
\int \frac{3}{2}\frac{\sqrt{g}\theta^{2}\phi^{2}}{N}\: dt\: d^{3}x \\
+ & \int 2N\sqrt{g}\grad_{i}\phi\grad^{i}\phi \: dt\: d^{3}x - 
\int 4N\sqrt{g}\phi\grad^{2}\phi\: dt\: d^{3}x \end{split}
\eeq
after some integration by parts. We notice that there might be a
possibility of ``completing some squares'' with terms involving K and those
involving $\theta$. We have,
\beq \label{ccfs} -N(trK)^{2} + N\kk -2\theta trK - 
\frac{3}{2}\frac{\theta^{2}}{N} \eeq
Let's try the combination,
\beq - N\biggl(trK + A\frac{\theta}{N}\biggr)^{2} + 
N\biggl(K_{ab} +B\frac{\theta g_{ab}}{N}\biggr)\biggl(K^{ab} + 
B\frac{\theta g^{ab}}{N}\biggr) \eeq
This gives us,
\beq - N(trK)^{2} - 2A\theta trK - A^{2}\frac{\theta^{2}}{N} + N\kk + 
2B\theta trK + 3B^{2}\frac{\theta^{2}}{N} \eeq
Comparing coefficients with equation (\ref{ccfs}) gives us,
\beq -2A + 2B = -2 \:\: \text{and} \: \: -A^{2} + 3B^{2} = -\frac{3}{2} \eeq
Solving here gives $ A = \frac{3}{2} $ and $ B = \frac{1}{2} $ and so we have,
\beq -N\biggl(trK + \frac{3}{2}\frac{\theta}{N}\biggr)^{2} + N\biggl(K_{ab} 
+ \frac{1}{2}\frac{\theta}{N}g_{ab}\biggr)\biggl(K^{ab} + 
\frac{1}{2}\frac{\theta}{N}g^{ab}\biggr)  \eeq
Finally, let us set  
\beq B_{ab} = \biggl(K_{ab} + \frac{\theta}{2N}g_{ab}\biggr) \eeq
Thus we get,
\beq - N\bigl(trB\bigr)^{2} + N\bb  \eeq
overall. Our full action is now,
\beq
\begin{split}
S = \;\; & 
\int N\sqrt{g}\phi^{2}\biggl(R - (trB)^{2} + \bb\biggr)\: dt\: d^{3}x \\
+ & \int 2N\sqrt{g}\grad_{i}\phi\grad^{i}\phi \: dt\: d^{3}x \\
- & \int 4N\sqrt{g}\phi\grad^{2}\phi \: dt\: d^{3}x \end{split}
\eeq
We are now in a $(3 + 1)$-dimensional form and so we would like to use the 
power of $\phi$ which is appropriate in $3$ dimensions. From the earlier 
discussion of conformal invariance in different numbers of dimensions we find 
that we should use $\psi = \phi^{1/2}$ (or $\psi^{2} = \phi$). This is no more
than a relabelling to make things look neater and there is no real change to 
the theory in this relabelling. We get, 
\beq
\begin{split}
S = \;\; & 
\int N\sqrt{g}\psi^{4}\biggl(R - (trB)^{2} + \bb\biggr)\: dt\: d^{3}x 
+  \int 8N\sqrt{g}\psi^{2}\grad_{i}\psi\grad^{i}\psi \: dt\: d^{3}x \\ & - 
\int 8N\sqrt{g}\psi^{2}\grad_{i}\psi\grad^{i}\psi \: dt\: d^{3}x 
- \int 8N\sqrt{g}\psi^{3}\grad^{2}\psi \: dt\: d^{3}x \end{split}
\eeq
Thus the action is
\beq \label{cgbk} S = \int N\sqrt{g}\psi^{4}\Biggl
(R - 8\frac{\grad^{2}\psi}{\psi} - (trB)^{2} + \bb\Biggr)\: dt\: d^{3}x \eeq
\\
This looks \emph{much} better! We notice too that 
$ \biggl(R - 8\frac{\grad^{2}\psi}{\psi}\biggr) $ is the conformally uniform 
3-D version of
$ \biggl(\,^{(n)}R - \frac{4(n -1)}{(n-2)}\frac{\Box{\psi}}{\psi}\, \biggr) $. 
In fact, if we start with the ADM (3+1) action, equation (\ref{admact}) and 
perform a conformal transformation
\beq
\begin{split} g_{ab} & \longrightarrow \psi^{4}g_{ab} \\ 
N & \longrightarrow \psi^{2}N \\
N_{i} & \longrightarrow \psi^{4}N_{i} \end{split}
\eeq
(which is simply $ g_{\alpha\beta} \longrightarrow \phi^{2}g_{\alpha\beta} $ ;
that is, a conformal transformation of the $4$-metric of the form equation(\ref{ct1})) we get precisely the 
action of equation (\ref{cgbk}).
\\ \\
\underline{Note}: We had $\theta$ in terms of $\phi$: $ \theta = 
-\frac{2}{\phi}\biggl(\dot{\phi} - \phi_{,i}N^{i}\biggr) $.
We may, of course, write it in terms of $\psi$: $ \theta = 
-\frac{4}{\psi}\biggl(\dot{\psi} - \psi_{,i}N^{i}\biggr) $.
We can also find a coordinate independent form for $\mathbf{B}$. This is
\beq \mathbf{B} = 
-\frac{1}{2}\psi^{-4}\mathbf{\pounds_{\underline{n}}}(\psi^{4}\mathbf{g}) \eeq
This is analogous to the expression
\beq \mathbf{K} = -\frac{1}{2}\mathbf{\pounds_{\underline{n}}}(\mathbf{g}) 
\eeq
for the extrinsic curvature $\mathbf{K}$ in general relativity.
\section{Jacobi Action}
Baerlein, Sharp and Wheeler \cite{bsw} constructed a Jacobi Action for general 
relativity.
Their action was,
\beq \label{bsw} S = \underline{+}\int 
d\lambda\int\sqrt{g}\sqrt{R}\sqrt{T_{\text{GR}}}d^{3}x \eeq
where
\beq T_{\text{GR}} = 
\biggl(g^{ac}g^{bd} - g^{ab}g^{cd}\biggr)\biggl(\frac{\partial 
g_{ab}}{\partial t} - (KN)_{ab}\biggr)\biggl(\frac{\partial 
g_{cd}}{\partial t} - (KN)_{cd}\biggr) \eeq
Variation with respect to $\frac{\partial g_{ab}}{\partial t}$ gives 
\beq \pi^{ab} = \sqrt{\frac{gR}{T}}\biggl(g^{ac}g^{bd} - 
g^{ab}g^{cd}\biggr)\biggl(\frac{\partial g_{cd}}{\partial t} - (KN)_{cd}\biggr)
\eeq
This expression is squared to give the Hamiltonian constraint. The variation 
with respect to $N_{a}$ gives the momentum constraint. The evolution equations
are found in the usual way. The equations found with the Jacobi action are 
those of general relativity if we identify $2N$ and $\sqrt{\frac{T}{R}}$. We 
want to construct the analogous case in this theory. Let us return to our (3+1)
action, equation (\ref{cgbk}),
\beq \mathcal{L} = N\sqrt{g}\psi^{4}\biggl(\rp - (trB)^{2} + \bb\biggr) \eeq
We can write the Lagrangian as 
\beq \mathcal{L} = \sqrt{g}\psi^{4}\Biggl[N\biggl(\rp\biggr) + 
\frac{1}{4N}\biggl(\beta^{ab}\beta_{ab} - (tr\beta)^{2}\biggr)\Biggr] \eeq
where $ \beta_{ab} = - 2NB_{ab} = \biggl(\frac{\partial g_{ab}}{\partial t} - 
(KN)_{ab} - \theta g_{ab}\biggr)$.
We now extremise with respect to $N$. This gives us,
\beq \label{Ntr} N = \underline{+}\frac{1}{2}\biggl(\beta^{ab}\beta_{ab} - 
(tr\beta)^{2}\biggr)^{\frac{1}{2}}\biggl(\rp\biggr)^{-\frac{1}{2}} \eeq
Substituting this back into the action gives us
\beq S = \underline{+}\int d\lambda\int\sqrt{g}\psi^{4}\sqrt{\rp}\sqrt{T}d^{3}x
\eeq
where $ T = \Bigl(\beta^{ab}\beta_{ab} - (tr\beta)^{2}\Bigr) $.
This is the conformal gravity version of the BSW action (\ref{bsw}).
\\ \\
Amazingly, this is \emph{precisely} the action that Barbour and \'{O} Murchadha
 found by starting with the BSW action and conformalising it under conformal 
transformations of the $3$-metric
\beq g_{ab} \longrightarrow \psi^{4}g_{ab} \eeq
The Jacobi action is manifestly 3-dimensional and its configuration space is 
naturally conformal superspace - the space of all 3-D Riemannian metrics modulo
diffeomorphisms and conformal rescalings. However, we found this action 
starting with a fully \emph{four-dimensional} theory!
\\ \\
Omitting the details (which are to be found in \cite{abfom}) the constraints 
are found to be
\begin{align} \label{c1} \pipi - g\psi^{8}\biggl(\rp\biggr) & = 0 \\
\label{c2} \grad_{b}\pi^{ab} & = 0 \\
\label{c3} tr\pi & = 0 \\
\label{c4} N\psi^{3}\biggl(R - 7\frac{\grad^{2}\psi}{\psi}\biggr) - 
\grad^{2}\biggl(N\psi^{3}\biggr) & = 0 \end{align}
The evolution equations are
\beq \frac{\partial g_{ab}}{\partial t} = Ng^{-\frac{1}{2}}\psi^{-4}\pi_{ab} +
(KN)_{ab} + \theta g_{ab} \eeq
and
\beq
\begin{split}
\frac{\partial\pi^{ab}}{\partial t} = & - \sqrt{g}N\psi^{4}\Biggl(R^{ab} 
- g^{ab}\biggl(\rp\biggr)\Biggr) - 
2Ng^{-\frac{1}{2}}\psi^{-4}\pi^{ac}\pi^{b}_{\;\; c} \\ & + 
\grad^{a}\grad^{b}\biggl(\sqrt{g}N\psi^{4}\biggr) - 
\sqrt{g}g^{ab}\grad^{2}(N\psi^{4}) + 4g^{ab}N\sqrt{g}\psi^{3}\grad^{2}\psi \\ &
 - 8\sqrt{g}\grad^{a}(N\psi^{3})\grad^{b}\psi + 
4g^{ab}\sqrt{g}\grad_{c}(N\psi^{3})\grad^{c}\psi \\ & - 
\grad_{c}N^{a}\pi^{bc} - \grad_{c}N^{b}\pi^{ac} + 
\grad_{c}\biggl(N^{c}\pi^{ab}\biggr) - \theta\pi^{ab} \end{split}
\eeq
\section{Conformally Related Solutions}
In conformal superspace conformally related metrics are equivalent. Thus, 
conformally related solutions of the theory must be physically equivalent and 
so it is crucial that we have a natural way to relate such solutions. Suppose 
we have one set of initial data $ (g_{ab},\pi^{ab}) $. These must satisfy the 
constraints (\ref{c2}) and (\ref{c3}). We solve the Hamiltonian constraint 
(\ref{c1}) for our ``conformal field'' $\psi$. Suppose now we start with a 
different pair $(h_{ab},p^{ab})$ where $h_{ab} = \alpha^{4} g_{ab}$ and 
$\rho^{ab} = \alpha^{-4}\pi^{ab}$. Our new initial data is conformally related
 to the original set of initial data. This is allowed as 
``transverse-traceless''-ness is conformally invariant and so our initial data 
constraints are satisfied. All we must do is solve the new Hamiltonian 
constraint for our new conformal field $\chi$ say. This constraint is now
\beq \rho^{ab}\rho_{ab} = h\chi^{8}\biggl(R_{h} - 
8\frac{\grad_{h}^{2}\chi}{\chi}\biggr) \eeq
The subscript $h$ on $R$ and $\grad$ is because we are now dealing with the new
metric $h_{ab}$. We now solve this for $\chi$. It can be shown that we must 
have $\chi = \frac{\psi}{\alpha}$. That is, $\psi$ is automatically transformed
when our initial data is transformed.
\\ \\
Now,
\beq  \chi^{4}h_{ab} = \frac{\psi^{4}}{\alpha^{4}}\alpha^{4}g_{ab} = 
\psi^{4}g_{ab} \eeq
and
\beq \chi^{-4}\rho^{ab} = \psi^{-4}\pi^{ab} \eeq
If we label these as $ \widetilde{\,g_{ab}} = \psi^{4}g_{ab} $ and 
$ \widetilde{\,\pi^{ab}} = \psi^{-4}\pi^{ab} $ than we can write our 
constraints as
\begin{align} \widetilde{\,\pi^{ab}}\widetilde{\,\pi_{ab}} - 
\widetilde{g}\widetilde{R} & = 0 \\
\widetilde{\,\grad_{_{b}}}\widetilde{\, \pi^{ab}} & = 0 \\
\label{sc1} \widetilde{tr\pi} & = 0 \\
\label{sc2} 
\widetilde{N}\widetilde{R} - \widetilde{\,\grad^{2}}\widetilde{N} & = 0 
\end{align}
All conformally related solutions are identical in this form (which is 
also the representation which most closely resembles general relativity). We 
shall call this the physical representation as it is the combination
$\psi^{4}g_{ab}$ which is the physical quantity. The momentum constraint 
is identical in the two theories.
\\ \\
This is the distinguished representation of \cite{abfom}. However, it was not 
shown there why or how this representation is the phsyical one. This is
 explicitly shown here.
\\ \\ 
The Hamiltonian constraint of general relativity on a maximal slice is 
identical to that here. The slicing equation in the physical representation 
looks just like the maximal slicing equation of general relativity. In this 
representation the evolution equations are
\beq \frac{\partial g_{ab}}{\partial t} = Ng^{-\frac{1}{2}}\pi_{ab} + (KN)_{ab}
\eeq
and
\beq
\begin{split} \frac{\partial\pi^{ab}}{\partial t} = & - \sqrt{g}N\Bigl(R^{ab} 
- g^{ab}R\Bigr) - 
2Ng^{-\frac{1}{2}}\pi^{ac}\pi^{b}_{\;\; c} \\ & + 
\grad^{a}\grad^{b}\biggl(\sqrt{g}N\biggr) - \sqrt{g}g^{ab}\grad^{2}N  \\
& - \grad_{c}N^{a}\pi^{bc} - \grad_{c}N^{b}\pi^{ac} + 
\grad_{c}\biggl(N^{c}\pi^{ab}\biggr) \end{split}
\eeq
These are exactly those of general relativity on a maximal slice. Thus, 
solutions of general relativity in maximal slicing gauge are also solutions 
here. There are of course solutions of general relativity which do not have a 
maximal slicing and these are not solutions of the conformal theory.
\\ \\
Consider again the full four-dimensional form of the theory. Suppose we have a
solution of the equations $g_{\alpha\beta}$ and $\phi$. If we perform a 
conformal transformation on this metric with conformal factor $\alpha$, say, 
the new metric $h_{\alpha\beta} = \alpha^{2}g_{\alpha\beta}$ must still be a 
solution. We find that the conformal factor this time is $\eta = 
\frac{\phi}{\alpha}$ and so we have $\phi^{2}g_{\alpha\beta} = 
\eta^{2}h_{\alpha\beta}$ which is yet another demonstration of the 
identification of conformally related solutions. In the physical representation
 the $4$-dimensional equations take the form of the Einstein equations in 
vacuum
\beq G^{\alpha\beta} = 0 \eeq
However, these are supplemented with the slicing conditions (\ref{sc1}) and 
(\ref{sc2}) thus giving the theory a distinguished slicing which sets it apart 
from general relativity.
\section{Topological Considerations}
So far we have not considered any implications which the topology of the 
manifold may have. In \cite{abfom} it was shown that if the manifold was 
compact without boundary then there would be frozen dynamics. That is, the only
 solution for the lapse would be $N \equiv 0$. The problem was resolved by 
introducing the volume into the action explicitly. With the necessary changes 
the constraints were found to be
\begin{align} \pipi - \frac{g\psi^{8}}{V(\psi)^{\frac{4}{3}}}\biggl(\rp\biggr) 
& = 0 \\
\grad_{b}\pi^{ab} & = 0 \\
tr\pi & = 0 \\
N\psi^{3}\biggl(R - 7\frac{\grad^{2}\psi}{\psi}\biggr) - 
\grad^{2}\biggl(N\psi^{3}\biggr) & = C\psi^{5} \end{align}
\\ \\
Of course, with the introduction of the volume term we have a change in the 
original four-dimensional action also. This becomes,
\beq S = \int \frac{\sqrt{-^{(4)}g}\phi^{2}\biggl(\,\fourr - 
\frac{6\Box\phi}{\phi}\,\biggr)}{V(\phi)^{\frac{2}{3}}}\; d^{4}x \eeq
We have an implicit $(3+1)$ split here because $V$ is a purely 
three-dimensional quantity.We vary with respect to $^{(4)}g_{0\alpha}$ and 
$^{(4)}g_{ij}$ separately. (We vary with respect to the lower index case as 
$^{(4)}g_{ij} = g_{ij}$ and so both the numerator and the denominator may be 
varied with respect to the spatial part of the metric.) The variations give
\beq \overline{G^{0\alpha}} = 0 \eeq
and
\beq N\sqrt{g}\phi^{2}\overline{G^{ij}} + \frac{2}{3}g^{ij}\sqrt{g}C\phi^{3}
= 0 \eeq
where
\beq C = 
 \int \frac{N\sqrt{g}\psi^{4}\biggl(\,\rp\,\biggr)}{V(\psi)}d^{3}x \eeq
arises, as usual, due to variation of the volume. As earlier, 
$\overline{G^{\alpha\beta}}$ is the Einstein tensor of the metric 
$\phi^{2}g_{\alpha\beta}$ and $\psi^{2} = \phi$. We have used the Hamiltonian 
constraint to simplify $C$.
\\ \\
We can combine the equations to get
\beq \label{star} \overline{G^{\alpha\beta}} + 
\frac{2}{3N}h^{\alpha\beta}C\phi = 0 \eeq
where $h^{\alpha\beta}$ is the induced $3$-metric. This has the form
\begin{gather}
\begin{Vmatrix} h^{00} & h^{0k} \\ \\  h^{i0} & h^{ik} 
\end{Vmatrix} \quad
= \quad \begin{Vmatrix} \;\;0\;\; & \;\;0\;\; \\ \\  \;\;0\;\; & g^{ik}
\end{Vmatrix}
\end{gather}
We may lower the indices using $g_{\alpha\beta}$ to get
\begin{gather}
\begin{Vmatrix} h_{00} & h_{0k} \\ \\  h_{i0} & h_{ik} 
\end{Vmatrix} \quad
= \quad \begin{Vmatrix} N^{s}N_{s} & N_{k} \\ \\  N_{i} & g_{ik}
\end{Vmatrix}
\end{gather}
\\ \\
In the physical representation equation (\ref{star}) becomes
\beq \label{ces} G^{\alpha\beta} + \frac{2}{3N}h^{\alpha\beta}C = 0 \eeq
where now $C = \Big<NR\Big>$.
\\ \\
Of these ten equations, the four $0\alpha$ equations are identical to those in
general relativity while the remaining six differ by the new term which arose 
due to the variation of the volume. This new term is both time dependent and 
position dependent and so behaves like a ``non-constant cosmological 
constant.'' It will undoubtedly lead to new features, particularly in 
cosmology. Some of these are discussed in \cite{abfom}. 
\\ \\
We must also do the variations with respect to $\phi$ and $\dot{\phi}$. The 
volume is independent of $\dot{\phi}$ and so this variation gives us exactly 
the same result as earlier, namely
\beq trB = 0 \eeq
However the volume is not independent of $\phi$ and so we will have a slight 
change. Varying with respect to $\phi$ gives us exactly what we found when we 
did the variation on the Jacobi form of the action (of course)
\beq N\psi^{3}\biggl(R - 7\frac{\grad^{2}\psi}{\psi}\biggr) - 
\grad^{2}\biggl(N\psi^{3}\biggr) = C\psi^{5}  \eeq
where 
\beq C = \int \frac{N\sqrt{g}\psi^{4}\biggl({\rp}\biggr)d^{3}x}{V(\psi)}\eeq
This becomes
\beq NR - \grad^{2}N = \Big<NR\Big> \eeq
in the physical representation.
\\ \\
Let us consider equation (\ref{ces}) again. Taking the trace gives us
\beq -^{(4)}R + 2\frac{\Big<NR\Big>}{N} = 0 \eeq
or
\beq N^{(4)}R = 2\Big<NR\Big> \eeq
If we average both sides of this equation we get
\beq \frac{\int N\sqrt{g}^{(4)}R d^{3}x}{\int \sqrt{g}d^{3}x} = 2\Big<NR\Big>
\eeq
It turns out that this equation is equivalent to the equation
\beq \frac{\partial tr\pi}{\partial t} = 0 \eeq
Of course, this is already known from the propagation of the $tr\pi$ 
constraint. Thus we have demonstrated that there is no inconsistency in the 
equations.
\section{Discussion}
The initial idea was to construct a theory with conformal superspace as its 
configuration space. These are $3$-dimensional ideas and it was not expected 
that such a clear $4$-dimensional picture would emerge. The clarity of the 
$4$-dimensional picture should allow easy comparison with aspects of general 
relativity which have traditionally been treated in the $4$-dimensional 
framework. Furthermore, this formulation may be useful in a path integral 
approach to quantisation.
\\ \\
The field equations of the theory are almost identical to Einstein's equations.
The differences are entirely due to the emergence of a preffered frame and it 
is this which breaks the explicit $4$-covariance of the theory. Of course, 
there will be a quite different cosmology not least due to the fact that since 
the volume does not change expansion is automatically ruled out along with 
anything explained by expansion (most notably the redshift). These are 
discussed in \cite{abfom} and so we will not delve any further into this here. 
\\ \\
There has been much work on other aspects of this theory. Among these are the 
Hamiltonian formulation including the constraint algebra and some 
Hamilton-Jacobi theory \cite{bk}, and coupling of the theory to matter 
\cite{abfom}. Cosmological considerations and quantisation are further issues 
which are currently being investigated. Furthermore, a number of related 
theories are also being actively investigated and will be the topics of future
 articles.
\\ \\
Whether or not conformal gravity proves to be a viable theory of gravity 
remains to be seen. Nonetheless, the quantisation of the theory may teach some 
valuable lessons with regard to a full quantum theory of gravity. This is in 
itself a worthwhile pursuit.
\section{Acknowledgements}
I wish to thank Niall \'{O} Murchadha for many discussions. These were 
invaluable from both technical and non-technical viewpoints. I would also like 
to thank Ed Anderson, Julian Barbour and Brendan Foster for discussions. This 
work was partially supported by Enterprise Ireland.

\end{document}